\documentclass[12pt,preprint]{aastex}
\usepackage[english]{babel}
\usepackage{graphicx}
\usepackage{xspace}
\usepackage{multicol}
\usepackage{todonotes}
\usepackage{fixme}
\newcommand{\jj}{\snname{2014J}}
\newcommand{\fe}{\snname{2011fe}}

\newcommand{\sn}{SN\xspace}
\newcommand{\sne}{SNe\xspace}
\newcommand{\snia}{SN~Ia\xspace}
\newcommand{\sneia}{SNe~Ia\xspace}
\newcommand{\snname}[1]{SN\,#1\xspace}

\def\lsim{\raise0.3ex\hbox{$<$}\kern-0.75em{\lower0.65ex\hbox{$\sim$}}}
\def\gsim{\raise0.3ex\hbox{$>$}\kern-0.75em{\lower0.65ex\hbox{$\sim$}}}

\shorttitle{Constraints on the progenitor system of SN~2014J from high-cadence early observations}
\shortauthors{Goobar et. al.}

\begin{document}


\title{Constraints on the origin of the first light from \jj}

%% Use \author, \altafilmark, and the \and command to format
%% author and altafilmarkiation information.

\author{%
A.~Goobar\altaffilmark{1},
M.~Kromer\altaffilmark{2},
R.~Siverd\altaffilmark{3},
K.~G.~Stassun\altaffilmark{3,4},
J.~Pepper\altaffilmark{3,5},
R.~Amanullah\altaffilmark{1},
M.~Kasliwal\altaffilmark{6},
J.~Sollerman\altaffilmark{2},
F.~Taddia\altaffilmark{2}
}

\altaffiltext{1}{The Oskar Klein Centre, Department of Physics, Stockholm University,
    SE 106 91 Stockholm, Sweden}
 \email{ariel@fysik.su.se}
\altaffiltext{2}{The Oskar Klein Centre, Department of Astronomy, Stockholm University,
    SE 106 91 Stockholm, Sweden}
\altaffiltext{3}{Vanderbilt University, Department of Physics \& Astronomy,
VU Station B 1807, Nashville, TN 37235, USA}
\altaffiltext{4}{Fisk University, Physics Department, 1000 17th Ave.~N., Nashville, TN 37208, USA}
\altaffiltext{5}{Lehigh University, Department of Physics, 413 Deming Lewis Lab, 16 Memorial Drive East
Bethlehem, PA  18015, USA}
\altaffiltext{6}{Observatories of the Carnegie Institution for Science, 813 Santa Barbara St, Pasadena CA 91101, USA}

\begin{abstract}
  We study the very early lightcurve of supernova 2014J (\jj) using
  the high-cadence broad-band imaging data obtained by the Kilodegree
  Extremely Little Telescope (KELT), which fortuitously observed
  M\,~82 around the time of the explosion, starting more than two months
prior to detection, 
  with up to 20 observations per night. These observations are
  complemented by observations in two narrow-band filters used in an
  H$\alpha$ survey of nearby galaxies by the intermediate Palomar
  Transient Factory (iPTF) that also captured the first days of the
  brightening of the \sn.  The evolution of the lightcurves is
  consistent with the expected signal from the cooling of shock heated
  material of large scale dimensions, $\gsim 1 R_{\odot}$. This could
  be due to heated material of the progenitor, a companion star or
  pre-existing circumstellar environment, e.g., in the form of an
  accretion disk. Structure seen in the lightcurves during the first
  days after explosion could also originate from radioactive material
  in the outer parts of an exploding white dwarf, as suggested from
  the early detection of gamma-rays. The model degeneracy translates into 
  a systematic uncertainty of $\pm 0.3$ days on the estimate of the first
  light from \jj.
\end{abstract}


\section{Introduction}
%Dedicated studies of SN~2014J will likely have
%unprecedented impact for studies of the physics of Type Ia supernovae,
%the content of the ISM in M~82 and improve our understanding of the circumstellar
%environment of the exploding star.
Type Ia supernovae (\sneia) rank among the sharpest tools to measure
the expansion history of the Universe. However, the ultimate limiting
factor in their use for precision cosmology rests on our ability to
control systematic uncertainties of astrophysical nature, most notably
due to extinction and a possible brightness evolution over cosmic time
\citep[see][for a recent review]{2011ARNPS..61..251G}.  Due to its
remarkably close distance, \jj in M\,~82 ($d\sim 3.5$ Mpc) presents us
with unique opportunities to study the environment and reddening of an
otherwise normal \snia \citep{2014ApJ...784L..12G,2014ApJ...788L..21A,
  2014ApJ...790....3K,2014arXiv1405.3970M,2014MNRAS.443.2887F,2014arXiv1408.2381B,2014arXiv1409.7066A},
and even the impact of the intervening dust on the polarization signal
\citep{2014arXiv1407.0136P,2014arXiv1407.0452K}.  Moreover, its
proximity allowed for the first time the direct detection of
gamma-rays from the $^{56}$Ni decay sequence
\citep{2014Natur.512..406C,2014arXiv1407.3061D} that power the optical
display of SNe~Ia. Contrary to expectations from most theoretical
models, the detection of characteristic lines of the $^{56}$Ni decay
arising 20 days past explosion seems to indicate the presence of
radioactive material close to the surface of the SN ejecta
\citep{2014arXiv1407.3061D}.

Prior to this event,
\fe in M\,~101 ($d\sim 6.4$ Mpc) was the best studied SN~Ia. In
particular, \citet{2011Natur.480..344N} and
\citet{2012ApJ...744L..17B} performed studies of the early lightcurve
to conclude that the explosion originated from a compact object, $R
\lsim 0.02 R_{\odot}$, i.e., the size of a white dwarf (WD),
% \mk{compatible sounds rather weak here. all previous limits were
% also compatible with WDs. the exciting thing with the Bloom work was
% that they excluded larger progenitors.}
as expected from models where \sneia result from a thermonuclear
explosion in a C/O WD \citep{1960ApJ...132..565H}.  The validity of
the tight limit has been questioned by \citet{2014ApJ...784...85P},
who suggested there could be a ``dark phase" of up to a few days
between the explosion  and the rise of the radioactively powered
  lightcurve. Due to the lack of well-sampled deep observations
prior to discovery, a possible early burst of light cannot be 
safely ruled out.  
% Allowing for a dark phase of \dots days they revised the limit for
% the progenitor size of \fe to $R \lsim 0.1 R_{\odot}$.
Further discussions on the early rise of \sneia have been provided by
\citet{2012ApJ...759...83P, 2013ApJ...769...67P} and
\citet{2014MNRAS.441..532D}.
% \mk{We might also want to mention the recent work by Dessart
% 2014MNRAS.441..532D
% \url{http://adsabs.harvard.edu/cgi-bin/bib_query?arXiv:1310.7747} He
% might be a potential referee. That's already one reason to cite the
% paper. But more important, these calculations are much more
% sophisticated than the Piro stuff and they do not show a t-squared
% law.}
From a spectroscopic analysis of \fe \citet{2014MNRAS.439.1959M}
derived a dark phase of $\sim 1$ day, resulting in a limit of $R \lsim
0.06 R_{\odot}$ for the progenitor size.
  
For \jj, \citet{2014ApJ...783L..24Z} reported January 14.75 UT ($\pm
0.21$ day) as the best estimate of the explosion date, based on
observations from the Katzman Automatic Imaging Telescope (KAIT) at Lick
Observatory  and amateur data from
K.~Itagaki\footnote{http://www.k-itagaki.jp/psn-m82.jpg}. They also
conclude that a fire-ball model \citep{1999AJ....118.2675R}, where the
luminosity is assumed to scale with the area of the expanding
photosphere at constant temperature and expansion velocity, $L \sim
t^2$, does not provide a good description to the KAIT data of \jj and
used instead a broken power-law expression to derive the time of first
light.  Similarly, \citet{2014ApJ...784L..12G} showed that the early
data of \jj from the intermediate Palomar Transient Factory (iPTF)
obtained with the 48-inch telescope (P48) in two narrow-band
H${\alpha}$ filters deviate from the simple $t^2$ law.

In this work, we investigate the possibility that the deviations from
a simple $t^2$ rise seen in the lightcurves of \jj originate from an
extra source of luminosity from either shock-heating of the ejecta,
interaction with circumstellar matter
% \mk{Probably better to say CSM instead of
%  accretion disc. The merger could lead to an extended envelope or an
%  accretion disc} 
or a companion star; or from radioactive 
material in the outer parts of the exploding star.  For this purpose, we
use the early iPTF narrow-band data from \citet{2014ApJ...784L..12G}
and the high-cadence observations from the Kilodegree Extremely Little
Telescope (KELT) from \citep{2014arXiv1411.4150S}

\section{Observations}
\begin{figure}[h]
%\epsscale{0.83}
\includegraphics[width=0.49\textwidth, trim = 1cm 0cm 0.1cm 0.5cm, clip=true]{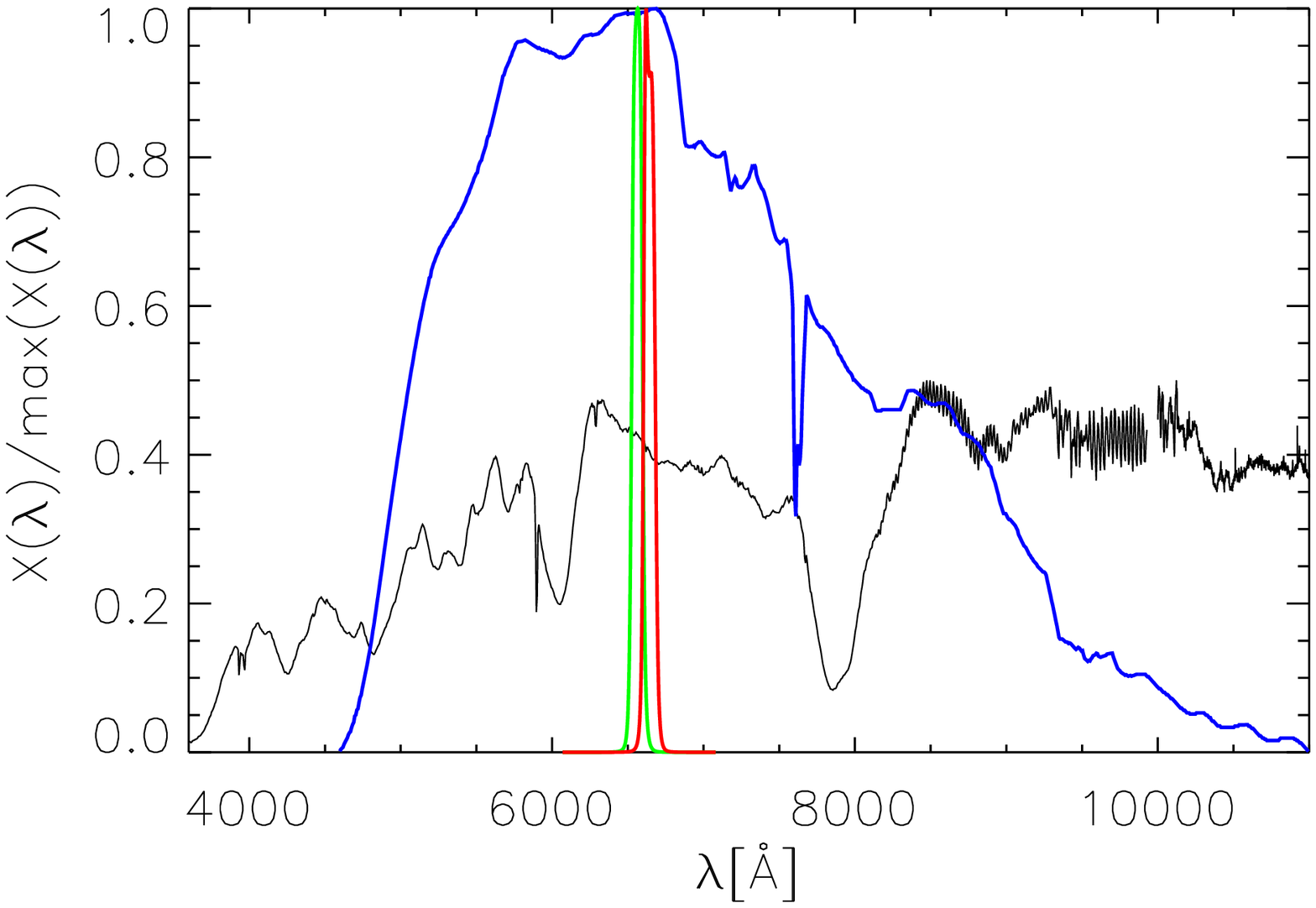}
\includegraphics[width=0.49\textwidth, trim = 1cm 0cm 0.1cm 0.5cm, clip=true]{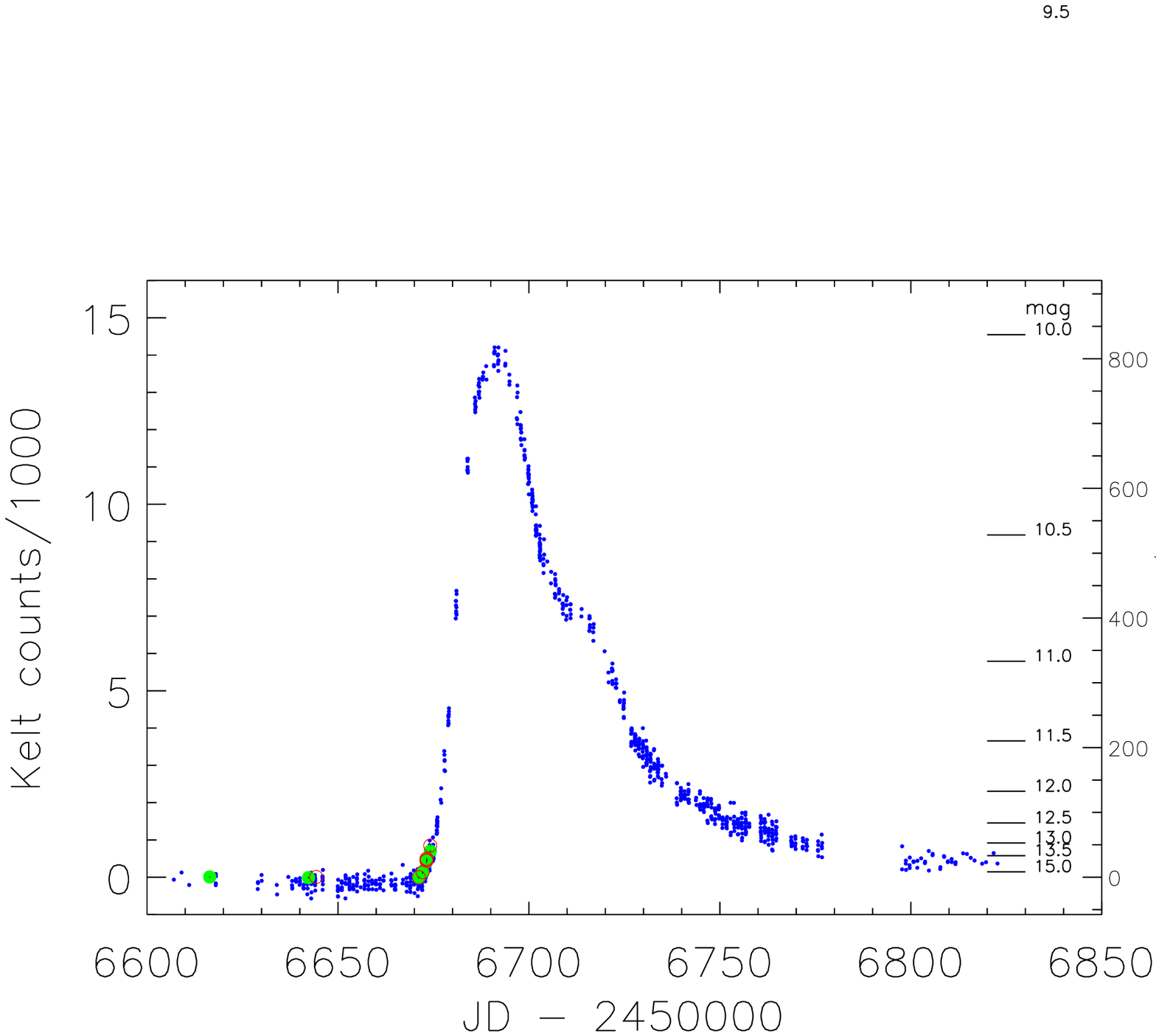}
\caption{{\em Left:} System transmission functions $X(\lambda)$ for
  the KELT (broad-band; blue) and iPTF (narrow filters; green/red)
  used for the early observations of \jj. Also shown in black is the
  earliest spectrum available, $F_\lambda$, from \citet{ATEL5786},
  about 7 days after the derived explosion date. {\em Right:} The KELT
  (blue) and iPTF (red, green) light curve points.  KELT covers epochs from about
  two months prior to explosion to day 140 after peak. Also shown are
  the natural Vega magnitudes for the calibrated fluxes.}
%
%\mk{Probably good to introduce two y-axis: Transmission [0,1]
%  on the left, and Scaled flux [arbitrary units] on the right?}}
 \label{fig:filter}
\end{figure}

%\todo[inline]{Describe data}
\label{observations}
The effective filter pass-bands (i.e., including the telescope optics, CCD sensitivity
and atmospheric throughput) used for the early observations of \jj
from iPTF \citep{2014ApJ...784L..12G} and KELT \citep{2014arXiv1411.4150S}
are shown in Fig.~\ref{fig:filter}, along with the \jj classification
spectrum \citep{ATEL5786} from January 22. 
The \sn exploded around full moon, when iPTF was conducting a
narrow-band survey of the Galactic plane and neighbouring galaxies
with the Palomar 48-inch telescope. Thus, the first set of iPTF
imaging observations were conducted with H$\alpha$ filters centred at
$6564$ \AA \, and $6638$ \AA \,, respectively.
Also shown in Fig.~\ref{fig:filter} is
the full (currently) available KELT lightcurve, starting about two months prior to
explosion and reaching to about day +140 after lightcurve maximum.
%\mk{Hmm, what you are showing are
%  transmission curves for KELT and your H$\alpha$ filters. I am not
%  sure if one would call that a wavelenght coverage of \jj.}  
%KELT is primarily an
%exoplanet transit survey \citep{2007PASP..119..923P}, using a small
%aperture telescope with a very wide FoV, $26^\circ \times 26^\circ$
%and a broad transmission function, $X(\lambda)$, cutting around
%$4600$\AA \, on the blue side and reaching into the near-IR,
%$\lambda\sim 1.1\mu$m.  It has a very coarse pixel scale, $\sim
%20{\arcsec}$, which makes accurate \sn photometry challenging (Siverd
%\etal, in prep.).

KELT is primarily an
exoplanet transit survey \citep{2007PASP..119..923P}, using a small
aperture telescope ($42$ mm) with a very wide FoV, $26^\circ \times 26^\circ$
and a broad transmission function, $X(\lambda)$, cutting around
$4600$ \AA \, on the blue side and reaching into the near-IR,
$\lambda\sim 1 \mu$m. The effective wavelength of the KELT pass-band 
(filter plus CCD response) is
$\lambda_{\rm eff} \approx 691$ nm. KELT has a very coarse pixel scale, $\sim
23{\arcsec}$, which makes accurate \sn photometry challenging \citep{2014arXiv1411.4150S}.
The KELT-North telescope obtained a total of 1869 individual exposures
of \jj that meet the quality control cuts. These exposures span 249 nights 
starting on 08 October 2013 and ending on 14 June 2014.

%(Julian Dates 2456573.96--2456822.69). 
Because the KELT-North telescope uses a German Equatorial mount,
observations taken before and after the local median crossing involve
a 180-degree flip of the FOV, which we designate West and East
exposures. In order to avoid systematics in the flux scale between the
West and East exposures, we restrict our analysis below to the West
exposures. Therefore the final KELT lightcurve includes 889 individual
exposures spanning 10 November 2013 to 14 June 2014 (Julian Dates
2456607.03 to 2456822.69).  
East and West are defined according to
  the sky position of the target, not by the orientation of the
 mount. In other words, the East exposures are acquired earlier in
 the night. % while the target is rising in the east. 
% Indeed, in line
%  with the referee's experience
 There is in general an offset (and a scaling) difference between the
 East and West photometry. Consequently, the East and West data are reduced
 independently and have a different flux scale. We found empirically
 that the West data have a better behaved PSF and smaller sky
 contamination, and therefore have overall better precision. Here we
 have opted to work with just the West orientation because of the
 better precision but most importantly to avoid systematics that could
 not be removed if attempting to also include the East
 data.
 %Indeed, one would normally opt to use the earlier data, however
 %fortunately in this case the time of first light is robustly
 %constrained with the west data, and the improved precision of the
 %west data permits the most stringent constraints on the progenitor as
 %described in the paper.  }

On nights with good weather where at least one measurement was
obtained, the number of measurements per night varies from 1 to 20.
Note that the KELT observing procedure employs a Moon avoidance
strategy, which leads to increased cadence on fields that are more
than 60 degrees away from the Moon. Fortuitously, this Moon avoidance
led to increased cadence on the M\,82 field around the time of the
explosion of \jj. Flux measurements from the individual KELT-North
frames are extracted using the difference-imaging aperture photometry
pipeline described in \citet{2014arXiv1411.4150S}, which is a modified
version of the standard KELT pipeline
\citep[see][]{2012ApJ...761..123S}. Briefly, the standard KELT
pipeline was modified to deal with the non-optimal placement of M\,82
near a corner of the KELT FOV, leading to poor extraction with the
standard pipeline. Instead, in the modified pipeline, the photometry
extraction was manually forced to a custom cutout region centered on
M\,82. Each photometric measurement is a differential
  measurement relative to a template image by means of image
  subtraction. The template image is the median of
  130 images obtained during the month prior to the explosion.

Typically, KELT achieves a photometric precision of 1\% or better on 
point sources brighter than $V\approx10$. In the case of \jj,
the KELT lightcurve achieved a photometric precision of 1.5\% 
on average at peak brightness, and 3\% on average at half peak brightness.

Photometry of the iPTF images was carried
out with the method used in \citet{2008A&A...486..375A}. PSF
photometry of the \sn fluxes of all epochs together with the
background host-galaxy light was fitted simultaneously.  The
degeneracy between the host and the SN was broken by fixing the SN
flux to zero for all epochs obtained in 2013.  Lightcurves were also
built of 240 stars in the fields, which were used for calibration as
well as for deriving systematic uncertainties between the epochs.  The
former was carried out by matching the field star photometry to 
SDSS DR10 $r'$ and $i'$ magnitudes.  The uncertainty for individual
epochs was found to be $0.03\,$mag and the uncertainty of the
calibration is good to $0.07\,$mag.  The photometry for the early rise
used in this analysis is summarized in Table~\ref{tab:data}.

\section{Early lightcurves of SNe~Ia}
%\mk{The title ``beyond radioactive heating'' is a bit misleading since
%  we are coming back to radioactive heating in Section 4.4. Probably
%  we should make it more general along the theme line light sources at
%  early times. In this case we could/should add another section on
%  surface radioactivity that could contribute. Could even mention the
%  CSM stuff of Levanon and then. This way we would discuss all the
%  processes that could contribute in this chapter and then turn to the
%  analysis in the next chapter.}

% The theoretical support for a power-law behaviour of the early
% radioactively powered rise time of \sneia has been questioned by
% \citet{2012ApJ...759...83P} and
% \citet{2013ApJ...769...67P,2014ApJ...784...85P}.  

Observationally, the early lightcurve shapes for \sneia follow, to a good
approximation, a power-law behaviour.  For \fe, hitherto the object with the
best-covered early-time lightcurve, \citet{2011Natur.480..344N} found
$L\propto t^\beta$, with $\beta=2.01\pm0.01$. Studies of stacked
lightcurves of a nearby \snia sample from the LOSS survey
\citep{2011MNRAS.416.2607G}, typically not quite as sensitive to the
very early phases, $t\lsim1$ day, found
$\beta=2.20^{+0.27}_{-0.19}$. Similarly, a lightcurve analysis of
SDSS-II supernovae at cosmological distances by
\citet{2010ApJ...712..350H} found $\beta=1.8^{+0.23}_{-0.18}$,
compatible with a similar study of SNLS \sneia, $\beta=1.8\pm 0.2$,
\citep{2006AJ....132.1707C}.  
Recently, \citet{2014arXiv1411.1064F} studied
a sample of 18 \sneia from the Palomar Transient Factory (PTF) and the La Silla-QUEST variability survey (LSQ) and found a mean value of $\beta=2.44 \pm 0.13$.

\citet{2011Natur.480..344N} and \citet{2012ApJ...759...83P} have shown
that such a power-law behaviour can in principle result from
radioactive heating of the supernova ejecta.  However, the exact shape
of the rise will depend on the particular gradients in velocity,
density and chemical composition within the ejecta \citep{2012ApJ...759...83P}. In
particular, estimating the SN explosion epoch by fitting a power law
to the early-time lightcurve might be misleading
\citep{2014ApJ...784...85P}. Explosion models that can explain the
observed properties of normal SNe~Ia like delayed detonations in
Chandrasekhar-mass WDs
\citep[e.g.][]{2009Natur.460..869K,2013MNRAS.436..333S,
  2014MNRAS.441..532D} or violent WD-WD mergers
\citep{2012ApJ...747L..10P}, predict that most $^{56}$Ni should be
buried in the central part of the ejecta and not reach to the
outermost layers.  As a consequence, it will take some time for the
radioactive energy to diffuse to the surface of the \sn ejecta and a
dark phase might occur between the onset of explosion and the time of
first light \citep{2013ApJ...769...67P, 2014ApJ...784...85P}.
%Moreover, additional energy sources may contribute at these earliest
%epochs.

\subsection{Cooling of shock-heated \sn ejecta}
\label{ejecta}
While the SN shock wave propagates through the progenitor star, it
deposits energy all over the ejecta. Due to efficient adiabatic
cooling this shock deposited energy becomes negligible (with respect
to radioactive heating) at $\sim$ 1 day after explosion for a WD progenitor. At very early
times, however, radiative diffusion from shock-heated outer layers of
the SN ejecta may contribute to the emergent luminosity. In the past,
several groups have investigated the cooling of shock heated SN ejecta
with analytical frameworks,
e.g. \citet{1992ApJ...394..599C,2010ApJ...708..598P,2010ApJ...708.1025K,2010ApJ...725..904N,2011ApJ...728...63R}. Assuming
spherically symmetric, radiation dominated, homologously expanding \sn
ejecta with constant opacity, the different approaches vary in their
treatment of radiative diffusion and opacity and their assumption of
the initial density and pressure profiles. However, recent work has
shown that the different approaches agree within a factor of $\sim$2
\citep{2011ApJ...728...63R}.

Here, we follow the approach in \citet{2011Natur.480..344N} and
\citet{2012ApJ...744L..17B} who use the model of
\citet{2011ApJ...728...63R} to calculate the early-time luminosity
($L^{\mathrm{E}}$ ) and temperature ($T^{\mathrm{E}}$) from shock
heated ejecta: 
%In particular we use Eqn. 1 of \citet{2012ApJ...744L..17B}
\begin{eqnarray}
L^{\mathrm{E}}(t)& =& 1.2 \cdot10^{40}R_{10}(E_{51})^{0.85}(M_\mathrm{c})^{-0.69}(\kappa_{0.2})^{-0.85}(f_\mathrm{p})^{-0.16}(t_\mathrm{d})^{-0.31}\, {\rm erg \, s}^{-1} \nonumber \\
T^{\mathrm{E}}_\mathrm{eff}(t) & = & 4.1\cdot10^{3} (R_{10})^{0.25}(E_{51})^{0.016}(M_\mathrm{c})^{0.03}(\kappa_{0.2})^{0.27}(f_\mathrm{p})^{-0.022}(t_\mathrm{d})^{-0.47}\, {\rm K},
\label{eq:ejecta}
\end{eqnarray}
where $R_{10}$ is the progenitor radius in units of $10^{10}$
cm, $E_{51}$ the explosion energy in units of $10^{51}$ erg,
$M_\mathrm{c}$ the progenitor mass in units of the Chandrasekhar mass ($1.38 M_{\odot}$),
$\kappa_{0.2}$ the opacity in units of $0.2$ cm$^{2}$g$^{-1}$,
$f_\mathrm{p}$ a form factor that depends on the density profile of
the progenitor star and $t_\mathrm{d}$ the time since explosion in
days.  In the following we adopt $E_{51}=1$, $M_\mathrm{c}=1$,
$\kappa_{0.2} =1$ and $f_\mathrm{p}=0.05$.

\subsection{Cooling of a shock-heated companion star}
\label{companion}
Similar to the situation discussed in the previous section, the SN
shock wave will heat the surface layers of a surviving companion star,
if present. This shock deposited energy may then contribute to
the early-time luminosity of the SN. This scenario was investigated by
\citet{2010ApJ...708.1025K}. From a multi-dimensional radiative
transfer simulation, Kasen showed that the emerging flux is strongly
angle dependent. For favourable viewing angles, i.e. those lying
within the hole that is carved out of the SN ejecta by interaction
between the ejecta and the companion star, shock heating of the
companion star leads to a significant flux excess at early times
compared to pure radioactive heating.  Hydrodynamical simulations of
interactions between the \sn ejecta and the companion star show that
the half opening angle of such ejecta holes is $\sim 30-40$ degrees
\citep[e.g.][]{2000ApJS..128..615M,2008A&A...489..943P}. In contrast,
from the opposite direction the optically thick ejecta are obscuring
the companion and no effect of the shock heating is present, as shown
in Fig.~2 of \citet{2010ApJ...708.1025K}.

\citet{2010ApJ...708.1025K} provides an analytic fit formula for the
evolution of the early-time bolometric luminosity and temperature, $L^\mathrm{C}$
and $T^\mathrm{C}$:
\begin{eqnarray} 
L^\mathrm{C}(t)& = & 10^{43}a_{13}(M_\mathrm{c})^{1/4}(v_9)^{7/4}(\kappa_\mathrm{e})^{-3/4}(t_\mathrm{d})^{-1/2}\, {\rm erg \, s^{-1}} \nonumber \\
T^\mathrm{C}_\mathrm{eff}(t)& =& 2.5\cdot10^4(a_{13})^{1/4}(\kappa_\mathrm{e})^{-35/36}(t_\mathrm{d})^{-37/72} \,{  \rm K}.
\label{eq:companion}
\end{eqnarray}
Here, $a_{13}$ is the separation distance between the \sn progenitor
and its companion in units of $10^{13}$ cm, $M_\mathrm{c}$ the
progenitor mass in units of the Chandrasekhar mass, $v_9$ the
expansion velocity of the SN ejecta in units of $10^9$ cm\,s$^{-1}$,
$\kappa_\mathrm{e}$ the electron scattering opacity in units of $0.2$
cm$^2$\,g$^{-1}$ and $t_\mathrm{d}$ the time since explosion in days. In
the following we adopt $M_\mathrm{c}=1$, $\kappa_\mathrm{e}=1$ and
$v_9=1$. Assuming that the companion fills its Roche lobe, its radius
can be constrained to be $\lesssim 0.5 a_{13}$.

\section{Analysis of the early lightcurves of \jj}

\citet{2014ApJ...783L..24Z} % and \citet{2014ApJ...784L..12G} 
pointed
out that the early broad-band lightcurve of \jj does not follow a unique power
law. Instead a broken power-law expression was used to fit the
data for the purpose of deriving the explosion date.  Apart from \jj,  there is today only one other SN known that
shows such a behaviour \citep[SN~2013dy,][]{2013ApJ...778L..15Z}. For
the bulk of observed \sneia the lightcurve rise is in good agreement
with a single power law
\citep{2006AJ....132.1707C,2010ApJ...712..350H,2011MNRAS.416.2607G}.

In what follows, we will try to reconcile the deviant behaviour of \jj
as being a combination of (at least) two effects: radioactive heating
from the inner ejecta and some extra component, e.g. from
shock-heating of the ejecta or a companion;  or surface radioactivity, 
described in Section \ref{shallow}.

%\mk{could back reference
%to the previous chapter where these processes have been explained.}

\subsection{The fitting procedure} 
Given the empirical evidence for a power-law behaviour for the
radioactive early lightcurve of a large body of \sneia, we assume
that the \sn flux through a system transmission function $X(\lambda)$
(see Fig.~\ref{fig:filter}) can be described by the sum of two
components, including one of the terms in Eqns (\ref{eq:ejecta}) or
(\ref{eq:companion}) and a power-law:
\begin{equation}
{\cal F}_X(t) = \cases{ C_X (t - t_0)^\beta + \int {{ \cal L}^{\delta}(\lambda,t-t_0) \over 4 d^2} X(\lambda) 10^{\left(-{A_\lambda \over 2.5 }\right)} d\lambda & $ t \ge t_0$ \cr
                                   0  & $t < t_0$,}
\label{eq:twocomp}
\end{equation} 
where $\delta$ is an index specifying the source of luminosity (e.g., 
heating of ejecta or companion star) and  
$A_\lambda$ corresponds to the adopted extinction law, where for \jj
we base our extinction corrections on the parameters in
\citet{2014ApJ...788L..21A}, and ${\cal L}^{\delta}(\lambda,t) $ is
the wavelength dependent luminosity:
\begin{equation}
{\cal L} ^\delta (\lambda,t) = L^{\delta}(t)  \cdot {B_\lambda(T^\delta_\mathrm{eff}(t)) \over \sigma_\mathrm{SB} \left[ T^{\delta}_\mathrm{eff} (t)\right]^4},
\label{eq:calL} 
\end{equation}
%\mk{Should introduce the meaning of $\delta$.}
where we  have assumed a black-body spectral energy distribution (SED) and used the
Stefan-Boltzmann law. 

Next, we use the above expressions to fit the
measured \sn lightcurves for the first four days after the explosion,
along with pre-explosion data to derive the explosion time, $t_0$ and
the physical parameters of the extra source of energy. For the case of
shock-heated ejecta, we fit the radius of the progenitor star. When
investigating the interaction with the companion, the parameter of
interest is the radius of the companion star, assuming Roche-lobe
filling.

%\mk{The statements above are written fully generally for contributions
%  from shock heating the ejecta or the companion, the following is
%  tailored to shock heating.}
It should be noted that the adopted phenomenological description in
Eq.~(\ref{eq:twocomp}) cannot be exact since, at any given time, the
two contributions will not be completely independent. E.g., the same parts
of the ejecta may contribute due to shock heating and radioactive
heating.  
However, in the limit of $t\rightarrow 0$ the shock heating
will dominate, while for $t\rightarrow \infty$ radioactive heating is
the main source of energy for the optical lightcurves. Similarly,
the fitted power-law component could absorb some of the contributions
of the extra source of power.

%\mk{When we observe a distant star, we see an average of the specific
%  intensity over the stellar disc in projection
%  $<I_\lambda>=\int_0^RI_\lambda(p)2\pi p dp/\pi
%  R^2=2\int_0^1I_\lambda(\mu)\mu d\mu = F_\lambda$, where $F_\lambda$ is the
%  astrophyscial flux. Since $B_\lambda$ is
%  isotropic, $<I_\lambda>=F_\lambda=B_\lambda$. Thus for a star of radius
%  $R$ at distance $d$, an observer sees the flux
%  $f_\lambda=F_\lambda\cdot \pi R^2/d^2$. Using Stefan-Boltzmann to
%  derive $R$ from $L$ and $T_\mathrm{eff}$ this results in the
%  expression without the $\pi$ in the denominator.}

We choose to fit the data in flux space (not magnitudes) to minimise
potential bias effects. This is important since for $t<t_0$ positive
as well as negative fluctuations should be treated
equally. Furthermore, the (mainly) Poisson noise is simple to
characterise in linear flux space.

\subsection{Verification of the method using observations of \fe}
\label{sec:sn11fe}
\begin{figure}[h]
%\epsscale{0.83}
\includegraphics[width=0.49\textwidth]{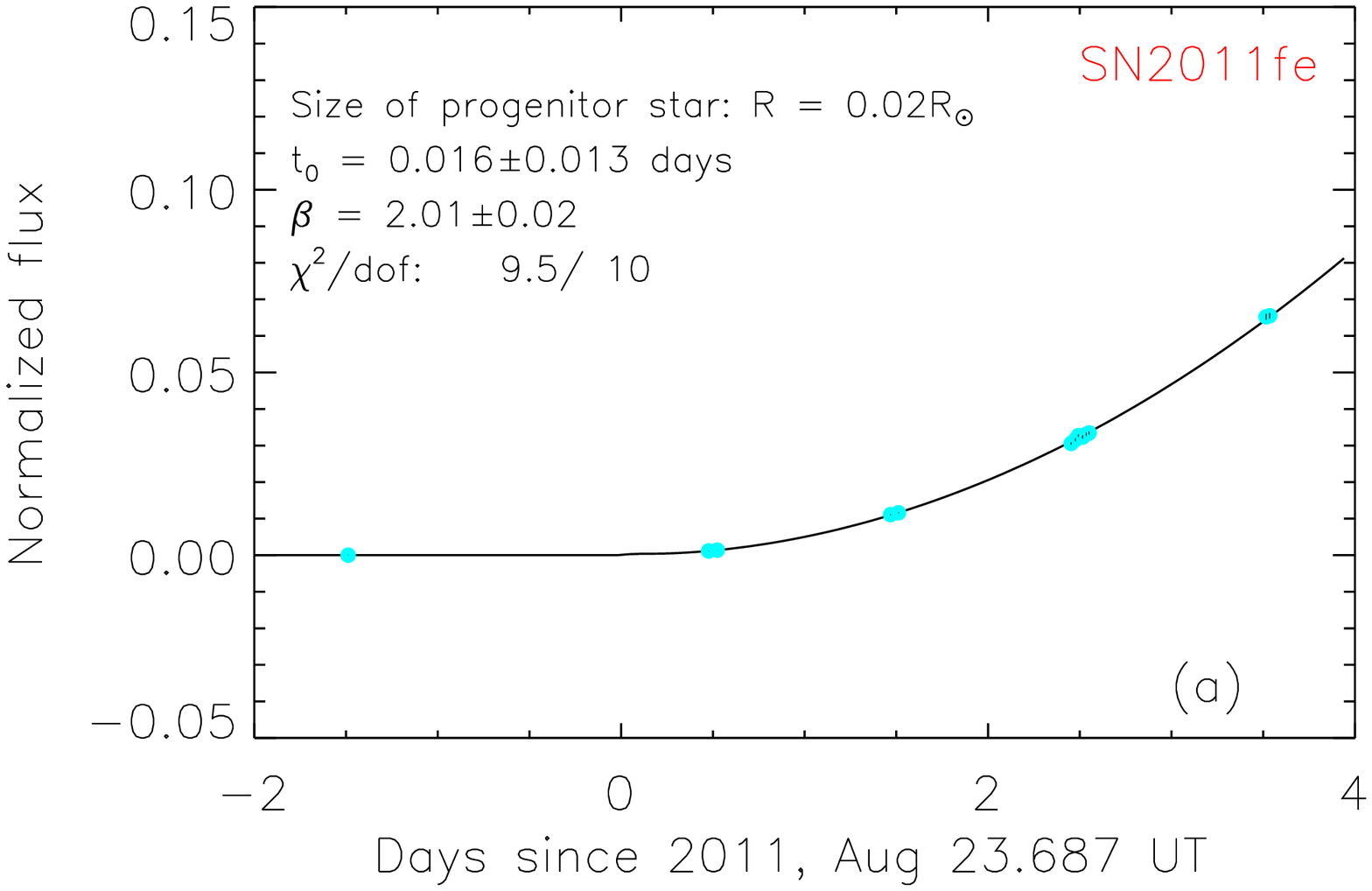}
\includegraphics[width=0.49\textwidth]{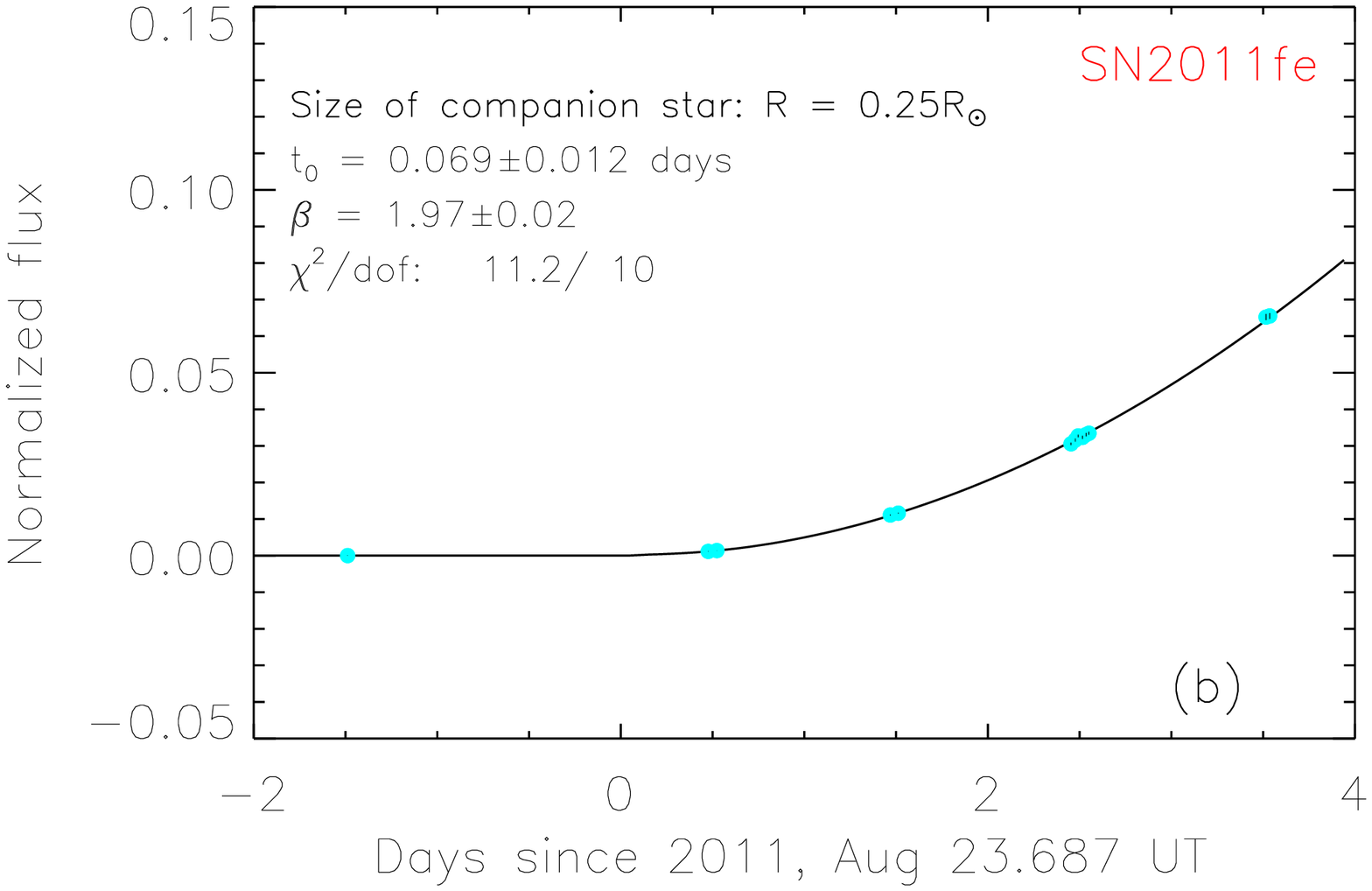}
\includegraphics[width=0.49\textwidth]{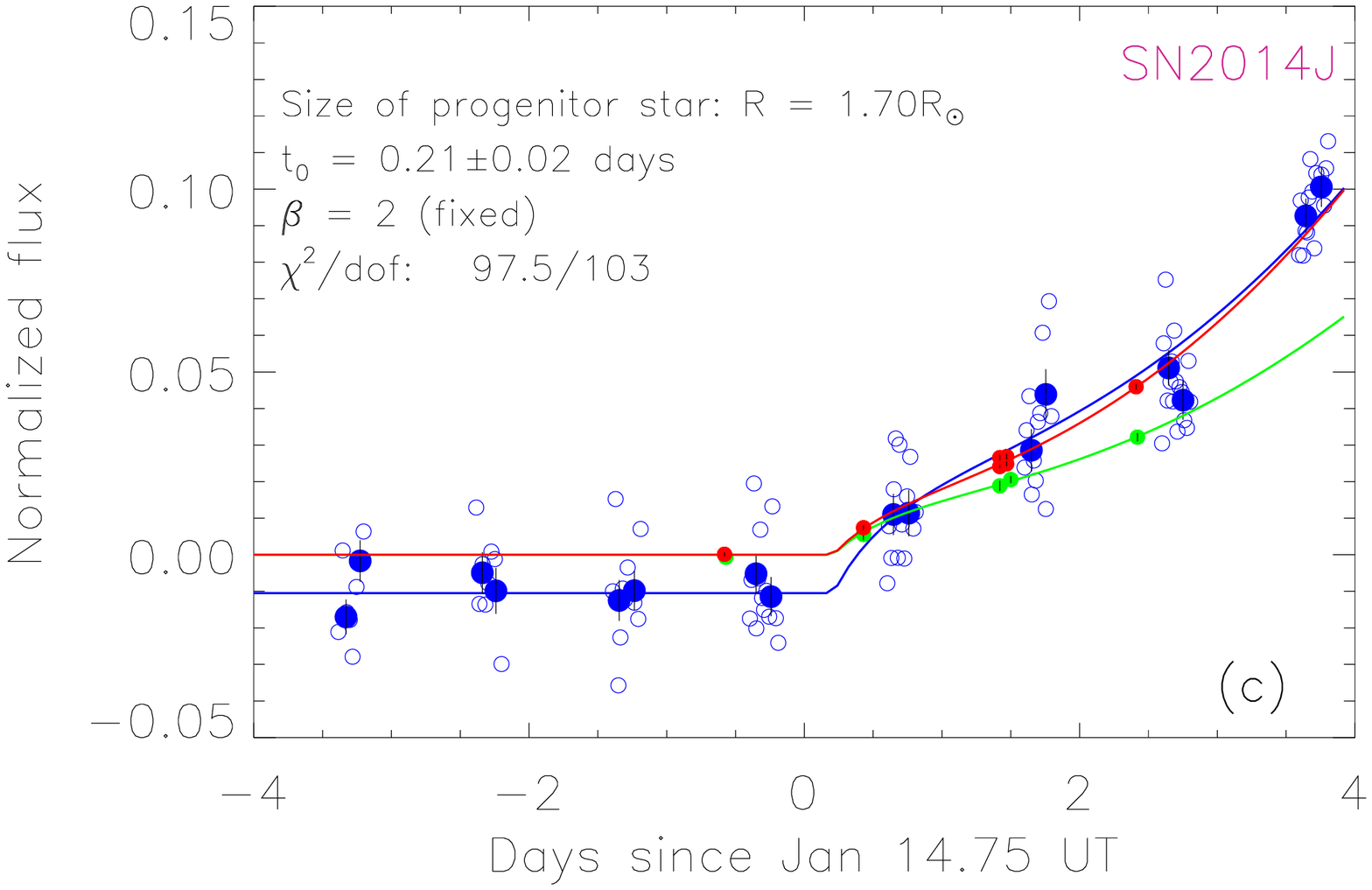}
\includegraphics[width=0.49\textwidth]{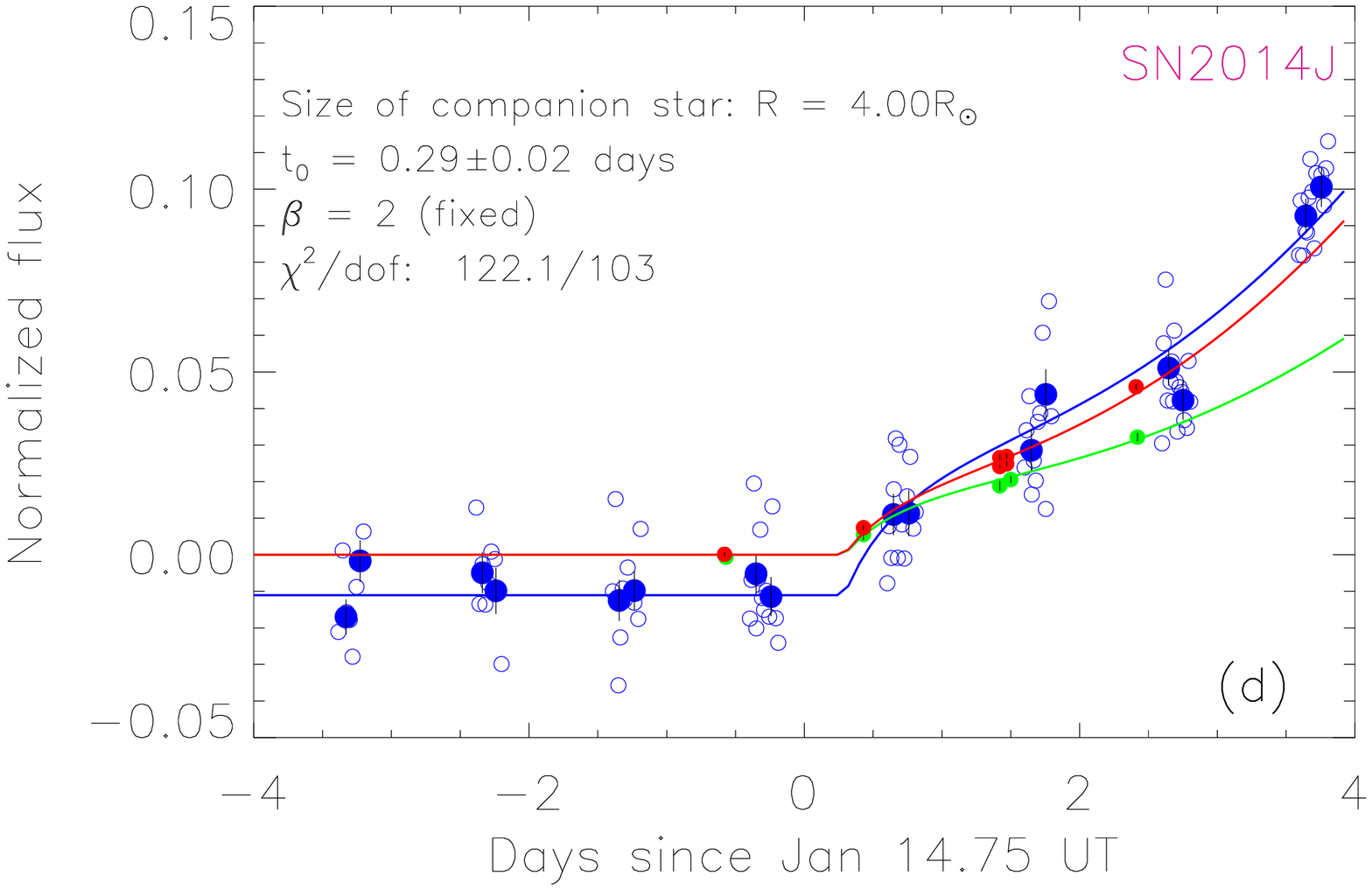}
%\plottwo{BetaEarly_fits_SN11fe0p02.eps}{BetaKasen_Early_fits_SN11fe0p25.eps}
%\plotone{BetaKasen_Early_fits_SN11fe0p25.eps}
\caption{Fits of early $g$-band lightcurve points and limits of \fe from 
\citet{2011Natur.480..344N} (above) and 
the broad-band KELT observations \citep{2014arXiv1411.4150S} and the two narrow-band $H_\alpha$ 
photometry of \jj from \citet{2014ApJ...784L..12G} (below). 
Open symbols show individual KELT photometry points, while the blue filled symbols
show the data combined into two bins for each night. Blue lines show
the fitted model.
Red and green bullets show the iPTF data for the two narrow-band filters shown 
in Fig.~\ref{fig:filter}. Similarly, the fitted models to those data points
are shown in same color.
Panels (a) and (c) show the fits where the lightcurves are assumed to have a contribution from
the heated progenitor ejecta. Panels (b) and (d) investigate the possibility
of heating of a companion star, following \citet{2010ApJ...708.1025K}. 
A small residual host galaxy contamination in the KELT photometry is fitted as a free (constant) parameter. }
 \label{fig:SN11fe}
\end{figure}

We first verify our method by revisiting the results on \fe based on the
$g$-band magnitudes reported by \citet{2011Natur.480..344N}, also
adopting their assumed distance to M~101 of 6.4 Mpc and negligible extinction $A_{\lambda}$.  For the epochs
with non-detections, we set the flux to zero and assign the
(symmetric) $1 \sigma$ error bar from the reported upper limits in
Table 1 of \citet{2011Natur.480..344N}.  The fit results are shown in
the two top panels of Fig.~\ref{fig:SN11fe}, where one can see that our
results are in good agreement with \citet{2011Natur.480..344N} and
\citet{2012ApJ...744L..17B}. In particular, we find that a combination
of a power-law model ($\beta=2$) with shock heating of a compact
progenitor of radius $R=0.02R_{\odot}$ provides an excellent fit, as
shown in Fig.~\ref{fig:SN11fe}(a). For the interaction with the
companion, as specified by the model of \citet{2010ApJ...708.1025K}
described in Section \ref{companion}, we find in
Fig.~\ref{fig:SN11fe}(b) that a companion star of size
$R\lsim0.25R_{\odot}$ provides a good fit to the observed $g$-band
lightcurve, i.e., slightly less constraining than the limits found by
\citet{2012ApJ...744L..17B}, $R < 0.1R_{\odot}$. The small
discrepancy is likely connected to the different analysis approaches, i.e.,
the use of flux space and two components in the fit.

\subsection{Fits of the combined iPTF/KELT data of \jj}
For the analysis of the early imaging data of \jj we use the same
methodology as for \fe in Sec.~\ref{sec:sn11fe}. For the system
transmissions of iPTF and KELT we use the functions $X(\lambda)$ shown
in Fig.~\ref{fig:filter}. Furthermore, we adopt the distance to M~82
from \citet{2009ApJS..183...67D}, $d=3.5$ Mpc, and the host galaxy
reddening of \jj derived by \citet{2014ApJ...788L..21A}, where we
opted to use their parameter solution for the Galactic type extinction
model of \citet{1999PASP..111...63F}, $R_V=1.4, E(B-V)_{\rm
  host}=1.37$ mag.  Exchanging to the multiple scattering model of
\citet{2008ApJ...686L.103G}, found to give a slightly better fit to the
full multi-wavelength lightcurve data from UV to the near-IR in
\citet{2014ApJ...788L..21A} and \citet{2014MNRAS.443.2887F}, does not
have a major impact on the outcome of the fits. We also include
Milky-Way reddening, $R_V=3.1$, $E(B-V)_{MW}=0.06$ mag. We have
assumed the dust causing the reddening is sufficiently far from the
site of the \sn that there is no significant change in the reddening
properties over time. The results of the fits for $\beta \equiv 2$ are
shown in the two bottom panels of Fig.~\ref{fig:SN11fe}. 

The lightcurves of the three filters are fitted simultaneously.
The number of degrees of freedom and the $\chi^2$ for the best fit 
are also indicated in the figure. Clearly, the fits are statistically
very sound and no systematic deviations are found for any of the data-sets.

Since there is an arbitrary flux offset between the
template image and the subsequent science images, this is 
accounted for in the lightcurve fitting through inclusion of a
constant offset term. Note that the offset
is in fact very small, comparable to the per-night binned photometric
precision (i.e., smaller than the per-point photometric precision) and
does not affect the results.

Fixing $\beta \equiv 2$, as observed in other SNe~Ia, we find that for the
scenario where the powering of the early rise of the lightcurve is a
combination of a power-law and shock heating of the ejecta, the radius
of the progenitor does {\em not} match that of a compact object for
\jj (at 90 \% confidence level), unlike the case for \fe. Instead, a radius in excess of the size
of the Sun is required. For the assumptions made, we find a best fit
for $R=1.7 R_{\odot}$.  \citet{2014arXiv1408.1375L} argue that such
large radii could be expected for WD mergers if an explosion does not
occur on the dynamical time scale of the merger, as in the violent
merger model \citep{2012ApJ...747L..10P}, but on the longer viscous or
thermal time scales of the merger remnant. In this case, angular
momentum conservation leads to the formation of an accretion disc or
extended envelope around the primary WD \citep{2012MNRAS.427..190S}
that extends to several solar radii.  Whether such a configuration was
present in the case of \jj cannot be definitely assessed from our
simple modeling. It is however interesting to note that
\citet{2014arXiv1407.3061D} speculate about the formation of a He
accretion belt to explain their detection of narrow gamma-ray lines
from $^{56}$Ni decay around maximum light. Detailed explosion models
will be required to address this question.
% One potential interpretation of this large radius is that it could
% correspond to the size of the disk-accretion region in the merging
% of two white-dwarfs, as in the model proposed by
% \citet{2014arXiv1408.1375L}. \todo{Markus, can you expand on this?}
% Connection with \citep{2014arXiv1407.3061D}....

When attributing the ``bump" in the early lightcurve to heating of a
companion star, we find that for the most optimal viewing angle
\citep[see Fig.~2 in][]{2010ApJ...708.1025K} a companion of radius
$4R_{\odot}$ is required. For less favoured viewing angles, much larger
companion stars would be needed to fit the observations. We also find that, if this lightcurve
modeling is assumed, the explosion occurred 7 hours (0.29 days) later than 
using the fit based on a broken power-law lightcurve in
\citet{2014ApJ...783L..24Z}. For the case of heated ejecta, our fits
suggest that the explosion took place 5 hours later than the best fit value of   
\citet{2014ApJ...783L..24Z}.
%\mk{The assumed explosion date is Jan
%  14.75 as shown in the axis and t0=0.29d in the caption is measured
%  relative to this? Then this amounts to an offset of $\propto7$
%  hours. Five hours are the case for the shock heated ejecta with
%  t0=0.21d. Probably we should discuss this for both cases
%  separately and say clearly that the estimated explosion time is
%  later by x hours than in Zheng (not just say within x hours).}
% \mk{make clear that
%  this rise time estimate depends on our model}. 

%\mk{Moved from the end of the section to here:} 

We note that the adopted formalism in Eq.~\ref{eq:twocomp} implies a negligible dark phase, since
 the power-law component as well as the extra heating
start at the same time, $t_0$. Although we could (in principle) add an extra free parameter in the fitting procedure to account for
a time offset between the two sources of light, the excellent goodness of fit in our results does not motivate
adding more degrees of freedom. We have also investigated the effect of
combining our data-set with the limits from \citet{2014ApJ...783L..24Z} and find that this
conclusion remains unchanged.

\begin{figure}[h]
%\epsscale{0.83}
\includegraphics[width=0.49\textwidth]{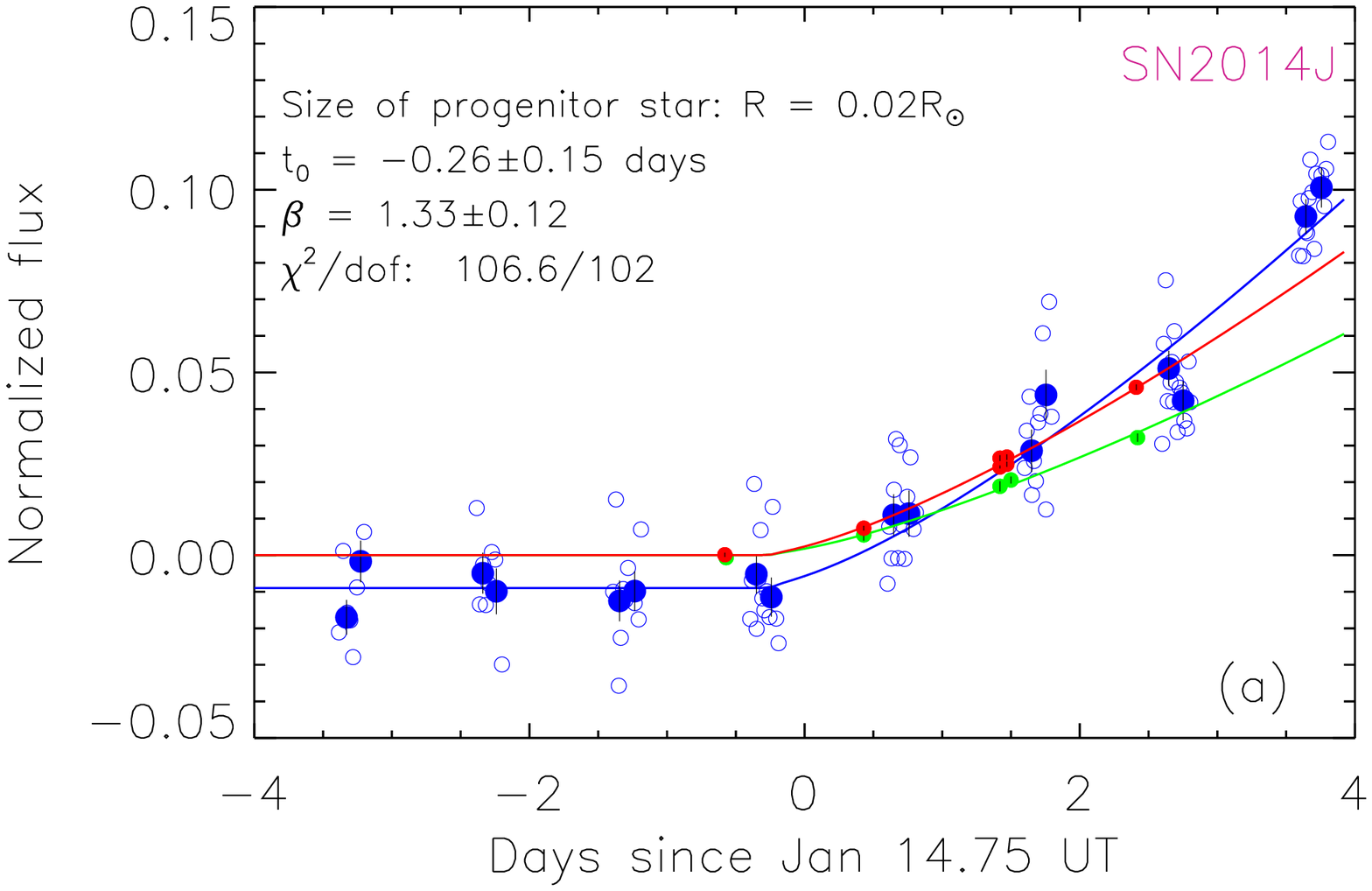}
\includegraphics[width=0.49\textwidth]{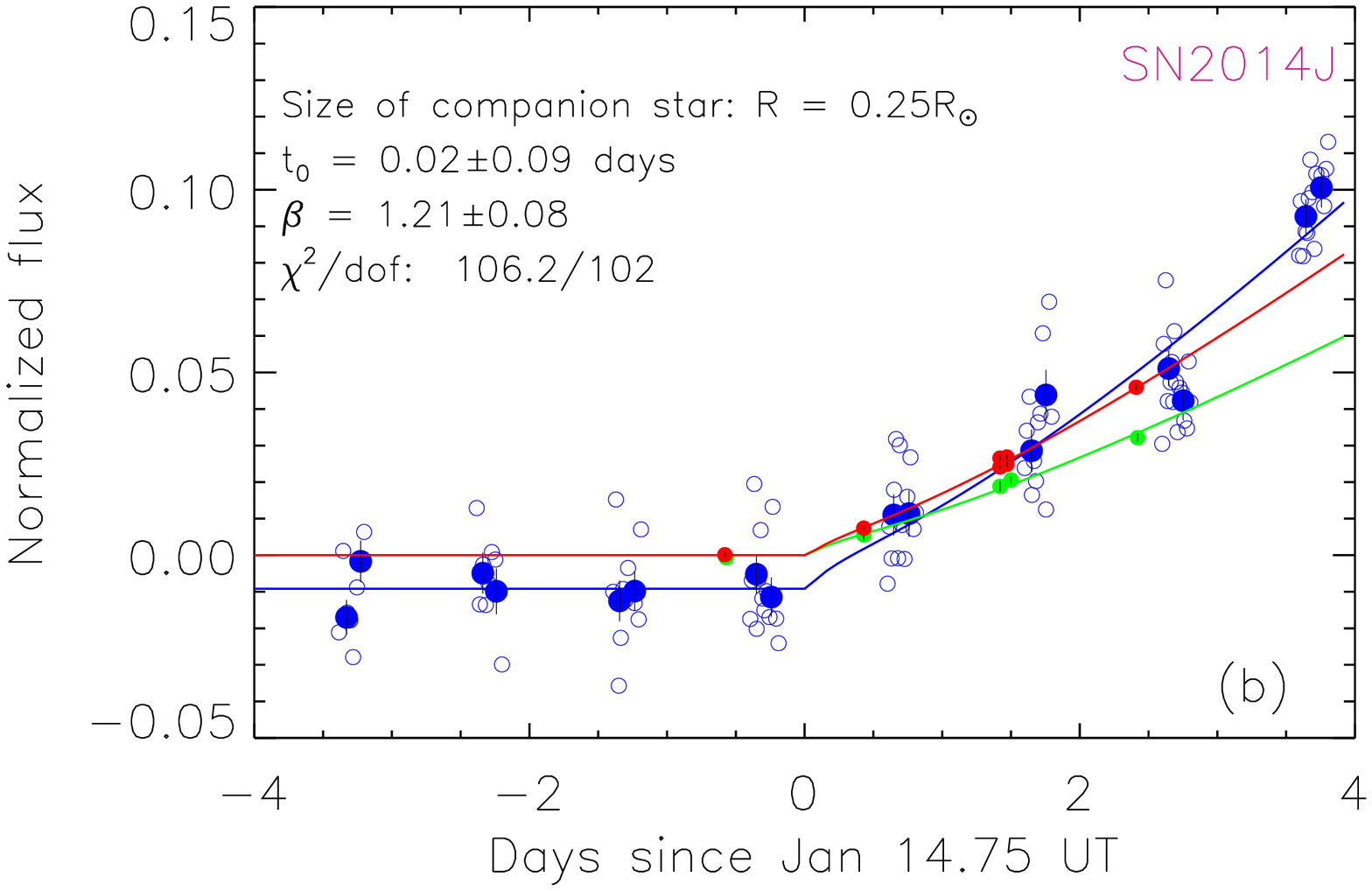}
\caption{Lightcurve fits of \jj fixing the size of the progenitor to be the
same as for \fe (a), and invoking the same size for an allowed companion 
star as for \fe  (b). Instead, the power-law parameter, $\beta$, is allowed to vary freely. Open symbols show individual KELT photometry points, while the blue filled symbols
show the data combined into two bins for each night. Blue lines show
the fitted model.
Red and green bullets show the iPTF data for the two narrow-band filters shown 
in Fig.~\ref{fig:filter}. Similarly, the fitted models to those data points
are shown in same color.
Good fits are
found, although for significantly flatter rise time, $\beta \sim 1.2-1.3$.}
 \label{fig:SN14JBeta}
\end{figure}

We note that it is still possible to get a formally acceptable fit for
a compact progenitor and/or a small companion but only by invoking a very
slow time dependence, $\beta=1.2-1.3$, as shown in
Fig.~\ref{fig:SN14JBeta}. According to \citet{2013ApJ...769...67P}
such a shallow rise is possible if direct heating from $^{56}$Ni is
dominating the luminosity indicating radioactive material right at
the surface of the SN ejecta. This is also required to explain the
detection of gamma-rays from $^{56}$Ni decay as reported by
\citet{2014arXiv1407.3061D}. The explosion time is slightly perturbed,
in the most extreme case by 6 hours (earlier) than in 
\citet{2014ApJ...783L..24Z}.

% \mk{In the context of Piro 2012 a power law
%   index $<2$ indicates larger velocity gradients than deposition
%   indices $\beta > \chi$, where $\beta$ is the power-law index of the
%   velocity profile and $\chi$ that for the 56Ni distribution. Piro \&
%   Nakar 2013 write ``that a rising light curve that is shallower than
%   $\sim t^2$ indicates that direct heating from 56Ni is dominating
%   ... [and] M56 can be approximated from the observations''. So this
%   would argue for surface radioactivity being
%   important/dominating. But I do not think that their statement is
%   very general. Within their model it will be true, but the model is
%   much simpler than reality. Piro \& Nakar 2014 write ``since most
%   explosion models predict a sharp decrease of X56 in the outermost
%   layers of the ejecta, the lightcurve is expected to rise
%   exponentially (due to diffusive tail contribution) at very early
%   times.'' } 

%\begin{figure}[h]
%%\epsscale{0.83}
%\includegraphics[width=0.49\textwidth]{NoBetaEarly_fits_1p70.eps}
%\includegraphics[width=0.49\textwidth]{NoBetax10Kasen_Early_fits_4p00.eps}
%\caption{}
% \label{fig:SN14JNoBeta}
%\end{figure}

\subsection{Surface radioactivity}%Radioactively powered lightcurve rise}
\label{shallow}
%\mk{This formulation appears in similar form also in the beginning of
%  the next chapter and in the introduction. While I would keep it in
%  the intro and the next Section, one could probably remove it
%  here. Instead, I would propose to describe physically how
%  radioactive material at different depths will affect the lightcurve
%  evolution. I.e.\ say that surface radioactivity means that the light
%  curve is sensitive to this material, while steeper gradients mean
%  that the light curve is sensitive to the diffusive tail. This is
%  already there in the second part of this paragraph. Maybe we start
%  with this and expand on it a little bit.}
%Concerns about the theoretical support for a universal $t^2$ behaviour
%have been raised by \citet{2012ApJ...759...83P} and subsequent
%papers. Perhaps the most sophisticated treatment of early-time SN Ia
%lightcurves to date has been performed by
%\citet{2014MNRAS.441..532D}. Generally, these simulations do not show a
%$t^2$ rise.
In the following we explore the possibility that radioactive material
was deposited close to the surface of the SN ejecta in more detail.
Surface radioactivity is actually predicted by various explosion
models, e.g.\ turbulent deflagrations that lead to completely mixed
explosion ejecta \citep[e.g.][]{2003Sci...299...77G,
  2014MNRAS.438.1762F} or double detonations in sub-Chandrasekhar-mass
WDs where an initial detonation in an accreted helium layer can lead
to radioactive isotopes close to the surface of the ejecta
\citep[e.g.][]{2010ApJ...719.1067K, 2011ApJ...734...38W}.
%Mention that anisotropies due to turbulent burning in deldet models
%could also lead to radioactive bubbles close to the surface.
How this material affects the early lightcurves in detail, depends
quite sensitively on its distribution and which isotopes are
produced. 

%\mk{I have reworded the following section. But we should be very clear
%  what we are actually doing. \citet{2013ApJ...769...67P} and
%  \citet{2014ApJ...784...85P} differ quite a bit in what they are
%  actually doing. While \citet{2013ApJ...769...67P} deals only with
%  the local contributions of direct heating and diffusive tail,
%  \citet{2014ApJ...784...85P} integrate the contribution of both
%  effects throughout the WD. I do not know which of these equations
%  you have coded up. Clearly \citet{2014ApJ...784...85P} is more
%  sophisticated but also more difficult. Also we should quote how we
%  derive $M_{56}$. This is model dependent and should be clarified.}

\citet{2013ApJ...769...67P} presented a simplified
analytical model to derive constraints on the $^{56}$Ni
distribution from the early bolometric lightcurve. Assuming that the
radioactively powered early lightcurves scale with the amount of
$^{56}$Ni ($M_{56}$) that has been exposed at time $t$, the luminosity
due to direct heating by $^{56}$Ni is $L^{\rm rad}(t) \approx
M_{56}(t) \cdot \epsilon(t)$, with:
\begin{equation}
\epsilon(t) =  \epsilon_{\rm Ni}e^{-t/t_{\rm Ni}} + \epsilon_{\rm Co} \left( e^{-t/t_{\rm Co}} - e^{-t/t_{\rm Ni}} \right),
 \label{eq:radioactivity}
\end{equation}
where $\epsilon_{\rm Ni, Co}$ and $t_{\rm Ni, Co}$ correspond to the
specific heating and decay time of $^{56}$Ni and $^{56}$Co,
respectively.  

\begin{figure}[h]
\centerline{\includegraphics[width=0.75\textwidth]{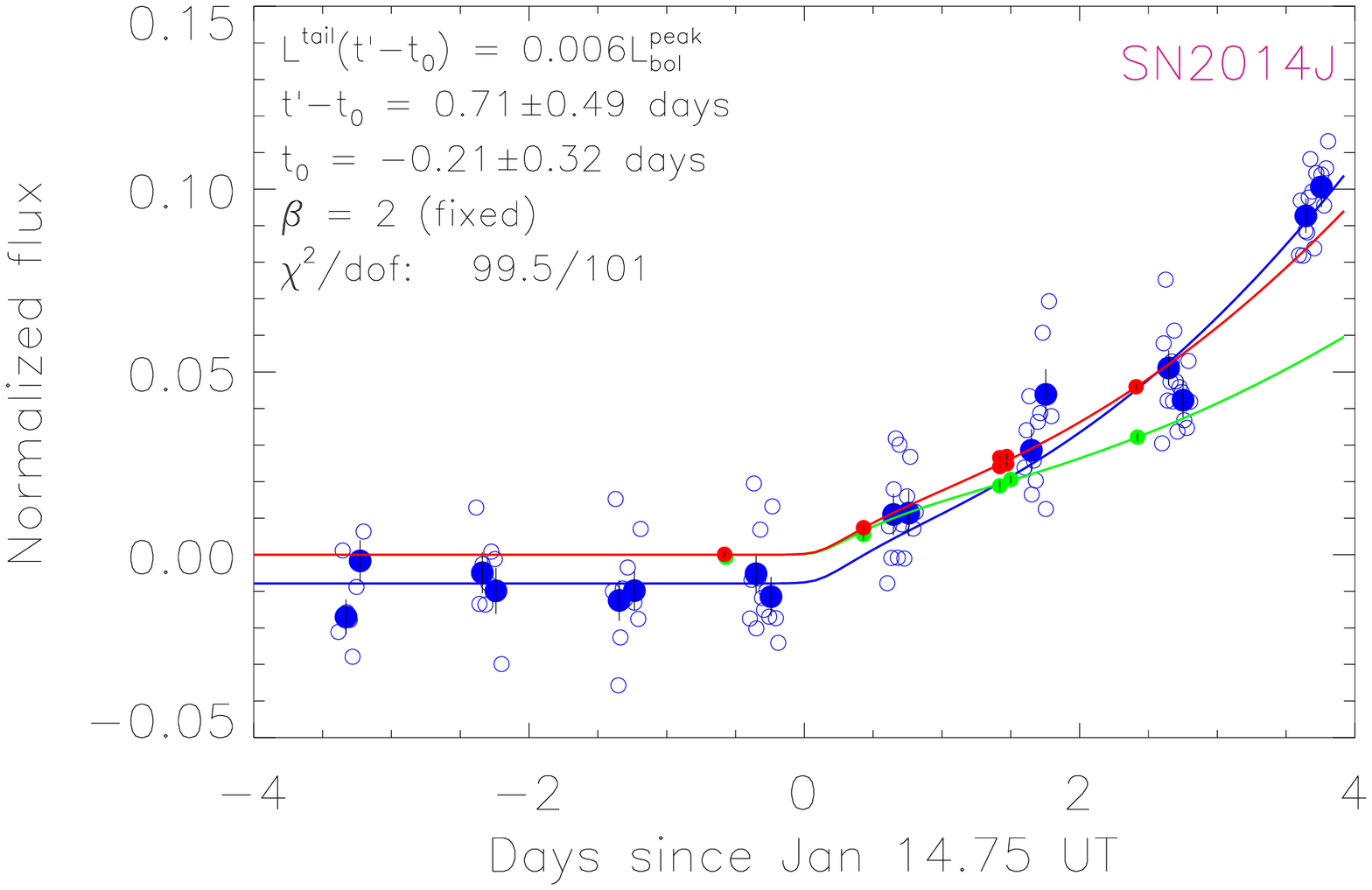}}
\caption{Lightcurve fits of \jj assuming early rise has contributions from
surface radioactivity \citet{2013ApJ...769...67P}, as described
in Section \ref{shallow}. 
 Open symbols show individual KELT photometry points, while the blue filled symbols
show the data combined into two bins for each night. Blue lines show
the fitted model.
Red and green bullets show the iPTF data for the two narrow-band filters shown 
in Fig.~\ref{fig:filter}. Similarly, the fitted models to those data points
are shown in same color.
}
 \label{fig:piro}
\end{figure}

For a steep gradient of radioactive material towards the inner parts
of the exploding star, the deposition of gamma-rays could be dominated
by deeper layers, with a fraction of photons escaping faster than the
average diffusion time.  \citet{2013ApJ...769...67P} call this a
``diffusive tail'' (schematically shown in their Fig.~2) and derive
the expected contribution to the bolometric lightcurve:
%L^{\rm tail} (\tau<\tau') \approx  L^{\rm rad}(\tau') {\epsilon(\tau) \over \epsilon(\tau')}  { {\rm erfc} \left( \tau'/\sqrt{2}\tau \right) \over {\rm erfc} \left(1/\sqrt{2} \right)},
\begin{equation}
L^{\rm tail} (t_0 \le t<t') = L^{\rm rad}(t') {\epsilon(t-t_0) \over \epsilon(t'-t_0)}  { {\rm erfc} \left( (t'-t_0)/\sqrt{2}(t-t_0) \right) \over {\rm erfc} \left(1/\sqrt{2} \right)},
\end{equation}
where % $\tau=t-t_0$,$\tau'=t'-t_0$ and
`erfc' is the complementary error function.  Next, we analyze the
early lightcurve data of \jj through fits of the theoretical models
described above making the crude assumption that the narrow and
broad-band iPTF and KELT fluxes are good proxies for the bolometric
luminosity, i.e., $L_{\rm bol}(t) \propto L_{X}(t)$, where $X$ correspond to the 
filter data available. Following a similar approach than in Sections \ref{ejecta}
and \ref{companion}, we consider the observed fluxes to be the sum of
two components:
\begin{equation}
{\cal F}_{X}(t) = C^{1}_{X}(t -t_0)^\beta + C^{2}_{X} L^{\rm tail} (t,t_0,t').
\label{eq:tail}
\end{equation}
Here we should clarify that this treatment differs from
  \citet{2013ApJ...769...67P}, since their model aims at explaining the full
  lightcurve evolution, without the additional power-law term. Our treatment
  is empirically motivated by the good $\chi^{2}$ of the fit, as shown in  Fig.~\ref{fig:piro},  which can be directly compared to the
  fits involving shock-heated material. 
  
 From the fitted luminosity, 
$L^{\rm tail}(t'-t_0)=6\cdot10^{-3} L^{\rm peak}_{\rm bol}$,
 and the assumption
that the bolometric peak brightness is as for \fe, $L^{\rm peak}_{\rm
  bol} =1.2 \cdot 10^{43}$ erg\,s$^{-1}$ \citep{2013A&A...554A..27P}, 
 we can estimate the amount of ``shallow'' ($\sim 0.1 M_{\odot}$ below surface)
$^{56}$Ni needed to power the lightcurve about 1 day after explosion,
$M_{56} \approx 1.1\cdot10^{-3} M_{\odot}$, corresponding to a mixing
mass fraction $X_{56}\approx 1.4 \cdot 10^{-2}$ when inserted in Eqn. 8
of \citet{2013ApJ...769...67P}.  We note that this falls short to match
the amount of surface $^{56}$Ni invoked by 
\citet{2014arXiv1407.3061D} in their modelling of the detection 
of gamma-rays from \jj. Given the possible degeneracy in our treatment, where the 
power-law component may absorb some of the contribution from radioactive surface
material, our estimate may be regarded as a lower limit.

\section{Conclusions}

We have analysed the early phase of the optical lightcurve following
the explosion of \jj in M\,82. The high-cadence KELT 
data, with up to twenty visits per night, starting two months prior to
the first detected light was used along with two sets of narrow-band
data from iPTF. The observations are suitable for studies of the 
first few days following the explosion. This data-set is unique among
published \sn lightcurves and can be compared to the previous
nearest ``normal'' \snia in modern time, \fe.

Unlike the case for \fe, a $t^2$-model together with cooling from
a compact progenitor does not provide the best match to the
observations. Instead, we find that the required size scale of the
heated material, either in the ejecta or a companion, is about a solar
radius or larger. As the possibility of having a single degenerate
model with a large companion has been ruled out by e.g., searches in
pre-explosion HST images \citep{2014ApJ...784L..12G,
  2014ApJ...790....3K}; non-detections at x-ray
\citep{2014ApJ...790...52M} and radio wavelengths
\citep{2014ApJ...792...38P}, other explanations to our findings are
needed. Following \citet{2014arXiv1408.1375L}, WD mergers could be one
possibility. If ignited on a viscous timescale rather than a dynamical
time scale, an extended CSM will form around the primary WD. Shock
heating of this material could be consistent with the observed early
lightcurve of \jj. Another possibility indicated by the shallow
lightcurve rise could be that surface radioactivity contributes to the
early lightcurve evolution. This was also invoked to explain the early
gamma-ray signal from $^{56}$Ni decay observed in \jj by
\citep{2014arXiv1407.3061D}, albeit in larger quantities than was
derived from our analysis of the optical lightcurve rise.  Based on
the available data, although with exquisite time sampling early on,
the degeneracies between the various explanations cannot be resolved,
due to insufficient signal-to-noise and color information.  The model
degeneracy translates into a systematic uncertainty of $\pm 0.3$ days
on the estimate of the first light from \jj.

We note that new well-sampled lightcurves are to be expected soon from
the ongoing KEPLER extragalactic
survey\footnote{http://www.astro.umd.edu/\~\,olling/KEGS\_K2.htm}. As
these will be carried on in an orchestrated fashion with surveys like
iPTF and KAIT, we can expect future discoveries to have a
richer spectroscopic and imaging coverage early on, e.g., in the UV
from Swift and HST. Methods to distinguish between the various
potential sources of luminosity in the early rise of Type Ia \sn
lightcurves, including temperatures and line velocities have been
proposed by \citet{2014ApJ...784...85P}. We also note that similar
attempts have been made to study the rise times of Ib/c \sne in
\citet{2014arXiv1408.4084T}.  Thus, the prospects of progress in this
area are good.

\acknowledgments 
AG and RA acknowledge support from the Swedish Research
Council and the Swedish Space Board.  
%MMK acknowledges
%support from the Hubble Fellowship and Carnegie-Princeton Fellowship.
%Observations made with the \spitzer telescope, the Nordic 
%Optical Telescope, operated by the Nordic Optical Telescope Scientific Association at the
%Observatorio del Roque de los Muchachos, La Palma, Spain, of the
%Instituto de Astrofisica de Canarias and the Mount Abu 1.2m Infrared 
%telescope, India. 

\begin{table}
\begin{center}
\begin{tabular}{ccc} \hline
% H656
\ & $H_\alpha^{656}$ & Zero-point = 24.77 \\ \hline
    UT Jan 2014 & flux & mag \\ \hline \hline
   14.18 &  -839.03 $\pm$   952.38 &    \   \\
   15.18 &  6933.76 $\pm$   653.08 &    15.17 $\pm$     0.08  \\
   16.17 & 23600.63 $\pm$   631.02 &    13.84 $\pm$     0.02  \\
   16.25 & 25762.67 $\pm$   654.94 &    13.74 $\pm$     0.02  \\
   17.17 & 40287.14 $\pm$   566.79 &    13.26 $\pm$     0.01  \\ \hline
%H663
\ & $H_\alpha^{663}$ & Zero-point = 24.83 \\ \hline
    UT Jan 2014 & flux & mag \\ \hline \hline
   14.17 &   131.28 $\pm$   970.02 &     \ \\
   15.18 &  7834.99 $\pm$   600.26 &    15.09 $\pm$     0.07  \\
   16.17 & 25596.23 $\pm$   694.16 &    13.81 $\pm$     0.02  \\
   16.17 & 28185.31 $\pm$   444.92 &    13.70 $\pm$     0.01  \\
   16.22 & 28485.04 $\pm$   830.09 &    13.69 $\pm$     0.03  \\
   16.22 & 26398.80 $\pm$   792.00 &    13.78 $\pm$     0.03  \\ 
   17.16 & 48755.88 $\pm$   713.46 &    13.11 $\pm$     0.01  \\ \hline
\ & KELT & Zero-point = 20.41 \\ \hline
    UT Jan 2014 & flux & mag \\ \hline \hline
   11.42 &  -241.71 $\pm$    67.48 &     \  \\ 
   11.52 &   -24.08 $\pm$    80.86 &     \  \\ 
   12.41 &   -70.10 $\pm$    79.43 &     \  \\ 
   12.51 &  -140.29 $\pm$    89.18 &     \  \\ 
   13.40 &  -177.58 $\pm$    79.28 &     \  \\ 
   13.52 &  -139.61 $\pm$    76.93 &    \  \\ 
   14.40 &   -73.86 $\pm$    75.38 &    \  \\ 
   14.51 &  -162.87 $\pm$    75.99 &    \  \\ 
   15.40 &   156.33 $\pm$    79.95 &    14.92 $\pm$     0.44  \\ 
   15.51 &   161.54 $\pm$    89.63 &    14.89 $\pm$     0.48  \\ 
   16.40 &   406.09 $\pm$    82.23 &    13.89 $\pm$     0.18  \\ 
   16.50 &   622.89 $\pm$    98.03 &    13.42 $\pm$     0.14  \\ 
   17.40 &   725.65 $\pm$    68.95 &    13.25 $\pm$     0.08  \\ 
   17.50 &   600.53 $\pm$    76.19 &    13.46 $\pm$     0.11  \\ 
   18.39 &  1316.60 $\pm$    66.64 &    12.61 $\pm$     0.04  \\ 
   18.51 &  1428.99 $\pm$    77.22 &    12.52 $\pm$     0.05  \\ \hline
\end{tabular}
\end{center}
\caption{Data used in this analysis. For KELT, only binned data is tabulated. The full data-set is presented in \citet{2014arXiv1411.4150S}}
\label{tab:data}
\end{table}

%{\it Facilities:} \facility{NOT}, \facility{Mount Abu Observatory}, \facility{P48},
%\facility{Keck}.

\bibliographystyle{apj}
%\bibliography{keltrise}

%\clearpage

\end{document}